\documentclass[aps,pra,showpacs,twocolumn,groupedaddress,amsmath,amssymb]{revtex4}
\usepackage{graphicx}

\begin{document}

\title{Spatially resolved photo ionization of ultracold atoms on an atom chip%
}
\author{S. Kraft}
\altaffiliation[Present address: ]{Van der Waals-Zeeman Instituut, Universiteit van Amsterdam, 
Valckenierstraat 65, 1018 XE Amsterdam, The Netherlands}
\email{S.Kraft@uva.nl}
\author{A. G\"{u}nther}
\author{J. Fort\'{a}gh}
\author{C. Zimmermann}
\homepage{http://www.pit.physik.uni-tuebingen.de/zimmermann/}
\affiliation{Physikalisches Institut der Universit\"{a}t
T\"{u}bingen, Auf der Morgenstelle 14, D-72076 T\"{u}bingen, Germany}
\date{\today}

\begin{abstract}
We report on photo ionization of ultracold magnetically trapped Rb
atoms on an atom chip. The atoms are trapped at 5 $\mu $K in a
strongly anisotropic trap. Through a hole in the chip with a
diameter of 150 $\mu $m two laser beams are focussed onto a
fraction of the atomic cloud. A first laser beam with a wavelength
of 778 nm excites the atoms via a two photon transition to the 5D
level. With a fiber laser at 1080 nm the excited atoms are photo
ionized. Ionization leads to depletion of the atomic density
distribution observed by absorption imaging. The resonant
ionization spectrum is reported. The setup used in this experiment
is not only suitable to investigate BEC ion mixtures but also
single atom detection on an atom chip.
\end{abstract}

\maketitle

\section{Introduction}

With microfabricated current conductors on a chip complex magnetic
field geometries can be constructed in which atoms can be trapped
and manipulated. Today, clouds of ultra cold atoms and
Bose-Einstein condensates are routinely trapped on such atom chips
and a variety of geometries have been demonstrated such as
waveguides, spatial and temporal beam splitters, double well
potentials, and periodic lattices (for a review see
\cite{Fortagh07}). It is now conceivable to develop atom chip
applications as for instance sensors for rotation, acceleration,
or gravitational forces gradients. The precision with which
condensates and ultra cold thermal clouds can be positioned at the
chip surface may also be exploited for surface probing and matter
wave microscopy \cite{Lin04,Wildermuth05,Guenther05a}. The distinct
suitability of atom chips for generating strongly anisotropic
trapping potentials also offers unique possibilities to study the
fundamental properties of the trapped quantum gas in the crossover
regime from three to one dimension. In very elongated chip traps
the phase coherence of a Bose-Einstein condensate breaks up giving
rise to phase fluctuations \cite{Esteve2006}. If, in addition, the gas is strongly
diluted and consists of only several ten atoms the regime of a
strongly interacting one dimensional Tonks-Girardeau gas may be
reached \cite{Paredes04, Kinoshita04}. Similar experiment with
fermionic atoms in the 1D regime \cite{Wonneberger06} may also be
possible \cite{Aubin06}. Such one dimensional quantum gases are of
great interest in fundamental many particle physics. Finally, if
methods can be found for carrying out experiments with only few or
even single atoms on a chip fascinating perspectives open up for
engineered quantum entanglement and quantum information
processing. Single atom detection on a chip is thus a promising
challenge with significant progress during the last year when
optical resonators have been successfully used to detect atoms by
optical spectroscopy \cite{Teper06,Haase06,Takamizawa06}.
Alternatively, atoms can be detected by ionization and subsequent
ion detection \cite{Campey06}. This approach is followed with this paper.

A second, still very young field of research is the investigation
of atom-ion-mixtures \cite{Cote02,Ciampini02,Massignan05}. In the
polarizing electric field of an ion the atoms are expected to form
a bubble with yet unknown properties. With many ions implanted in
the dilute quantum gas a system appears which has not been studied
even theoretically. A planar chip trap for ions has already been
demonstrated \cite{Seidelin06} and its integration on an atom chip
seems feasible. In such combined traps for atoms and ions the
magnetic field generating conductors trapping the atoms could
simultaneously be used as electrodes for generating the electric
field for trapping the ions. Together with the highly flexible
geometries that can be realized on atom chips, atom ion systems
offer fascinating new opportunities for constructing and
investigating novel types of quantum systems.

In this paper we demonstrate the controlled production of cold
ions by photo ionization of ultracold neutral atoms on an
atomchip. The ions are generated in a small spatially resolved
region given by the focus of the ionizing laser beams. The
generated ions with kinetic energies corresponding to temperatures
of only a few millikelvin may in principle be trapped in planar
ion traps and subsequently cooled by thermalization with a
surrounding atomic gas. Furthermore, spatially resolved photo
ionization with close to 100\% efficiency allows for single atom
detection on a chip. With a suitable ion optics the ions can be
extracted from the chip and detected with an electron multiplier
(CEM). Such a detector has recently been described in
\cite{Stibor07}.

In the next two chapters we describe a scheme for fast spatially resolved
photo ionization and its implementation on the chip. Chapter three and four
presents the experimental observations and concludes the paper.

\section{Ionization scheme}

In our setup $^{87}$Rb atoms are prepared in the 5S ground state.
\begin{figure}[tbp]
\includegraphics{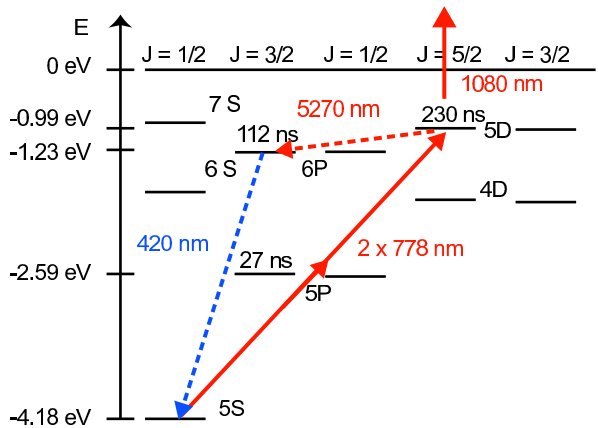}
\caption{(Color online) Ionization scheme. $^{87}$Rb atoms in the 5S ground state
are resonantly excited by two photon absorption into the 5D level (2 $\times$ 778 nm).
Before they decay back into the ground state the excited atoms are
rapidly ionized with a second laser (1080 nm).} \label{steps}
\end{figure}
The   atoms are ionized by a resonant two photon transition into the 5D excited state and subsequent excitation
to the continuum with a second laser (see Fig. \ref{steps}). The
two photon transition is driven by a grating stabilized continuous
wave single mode laser diode with a wavelength of 778.1066 nm. The
transition strength is enhanced by the intermediate 5P level which
is off resonant to the laser by only 2 nm. The 5D level with a
lifetime of 220 ns gives rise to a resonance line width of
approximately 500 kHz \cite {Nez93}. The rate of the two photon
transition depends quadratically on the light intensity which
leads to an improved spatial resolution and to a reduction of
unwanted stray light effects. This is important since we aim at a
fast ionization of the directly illuminated atoms on a time scale
of several microseconds while all other atoms should not be
affected by the light on a time scale of seconds, which is set by the lifetime of the condensate. For a two photon
transition stray light must thus only be suppressed to a level of
0.1\% in contrast to a single photon transition which requires a
suppression stronger by three orders of magnitude.

Furthermore, a hyperfine resolving optical excitation scheme can
be exploited to improve the spatial resolution of the ionization:
for atoms prepared in the 5S, F=2, m$_{F}$=2 hyperfine state and a
laser tuned in resonance with the F=1, m$_{F}$=1 hyperfine state
ionization only occurs if the two states are coupled by an
additional radio frequency. Due to the Zeeman effect resonant
coupling only occurs at a distinct magnetic field that depends on
the detuning of the radio frequency. In a magnetic field gradient
ionization then takes place only at a well defined position which,
in addition, can be varied by simply tuning the radio frequency.
At atom chips large field gradients can be easily constructed and
a resolution of below 100 nm appears feasible. Thus, there is no
principle obstacle for resolving an atomic distribution on a scale
that is comparable to the healing length of a condensate or the
inter atomic distance in a Tonks gas. This scheme also further
suppresses stray light effects and it should be possible to detect
atoms in a gas without affecting their neighbors.

For ionizing from the 5D level we use a continuous wave fiber laser near 1080nm with a Gaussian single mode beam profile
and a spectral width of smaller than 1 nm. Photons at this
wavelength are sufficiently energetic to bridge the binding energy
of the 5D level of 0.99 eV corresponding to a wavelength of 1250
nm. In principle the photon energy of the diode laser near 778 nm
is sufficient to ionize the excited atoms, however, efficient
ionization requires higher power than available with the diode
laser.

In order to reach a high ionization efficiency the ionization rate
from the 5D state to the continuum should be at least on the order
of the spontaneous decay rate to the ground state. The ionization cross section from the 5D level is reported to be
25 Mb \cite{Duncan01} for light near 1250 nm and drops to $\sigma
=17.5$ Mb for the shorter wavelength of the fiber laser near
$\lambda =$1080 nm \cite{Duncan01}. For the moderate intensities
discussed here, ionization is well described by a simple rate
model which relates the intensity $I$ and the ionization rate $R$
by:
\begin{equation}
I=\frac{R_{1}hc}{\lambda \sigma }
\end{equation}

Consequently, an ionization rate of $R_{1}=1/(220$ ns) requires an
intensity of $ I_\mathrm{C} = 4.6\cdot 10^{8}$ W / m$^{2}$. The
laser beams in our experiment are focussed to a $1/e^2$ radius of
$w_0 = 30 \mu$m. The ionization shall take place within this
radius and hence the intensity $ I_\mathrm{C}$ has to be reached
at $w_0$. For a Gaussian beam with this properties this results in a total
power of about 4.8 W which is easily available with a low cost
commercial fiber laser.

The two photon transition can also be treated in a rate model,
given that its rate $R_{2}$ is significantly slower than
ionization rate from the 5D level \cite{Ackerhalt76}. Then,
$R_{2}=3\cdot 10^{-4}(m^{2}/W)^{2}s^{-1}\cdot I^{2}$
\cite{Grove95}. For a beam radius of  $ w_0 = 30 \mu$m a laser
power of only 6 mW already results in an excitation rate of
$1/(100 \mu$s). 

%\subsection{Dipole potentials}

The laser light not only excites the atoms, it also gives rise to
light induced dipole potentials which have a strong impact on the
atomic motion. The most important contribution in Rb is related to
the resonance between the 5S ground state and the exited
5P$_{3/2}$ level (D2-line). The diode laser is 2 nm blue detuned
relative to this transition and leads to a repulsive potential
barrier with a hight of 5 $\mu $K  for the above parameters. In
contrast, the fiber laser is red detuned by 300 nm and generates
an attractive potential of 10 $\mu $K. Together both effects
almost cancel with a residual attractive net potential which drags
the atoms into the ionization volume.

\begin{figure}[tbp]
\includegraphics[width=8cm]{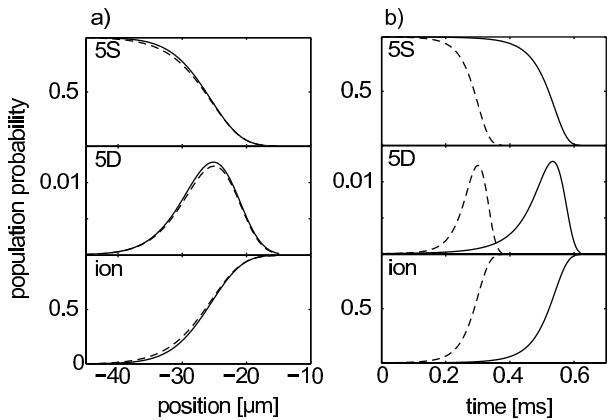}
\caption{Simulation of the ionization. The population probability of the different
states is plotted for different positions (a) and times (b) as the atom is
draged into the laser beams. The beam radius of the two lasers is set to be $w_0=30 \mu$m (solid lines). Increasing the waist of the fiber laser to $60 \mu$m (dashed lines) results in an earlier and faster ionization.}
\label{levels}
\end{figure}

Figure \ref{levels}
shows a simulation of the ionization process including the center
of mass motion of the atom in the optical dipole potential. The
laser powers are set to 6 mW and 5 W for the diode laser and the
fiber laser, respectively. In (a) the occupation of the involved
states is plotted versus the position of the atom which starts
with a small but final velocity (14 $\mu $m/s) at a distance of 60 $\mu$m from
the focus of the lasers (located at $x=0$). At a distance of
approximately 40 $\mu $m from the center of the laser focus the
probability for excitation in the 5D state grows significantly.
The solid line shows the simulation with a beam $1/e^2$ radius of
$w_0 = 30 \mu$m for both lasers. If the beam radius of the fiber
laser is expanded to 60 $\mu $m (dashed line) the results in (a)
remain almost unchanged. Since for this set of parameters the two
photon transition forms a bottleneck for ionization the internal
dynamics starts not until the atom reaches the light of the diode
laser. However, widening the focus of the fiber laser results in
earlier and faster ionization (dashed lines in Fig.2 (b) due to the
attractive dipole potential of the fiber laser which accelerates
the atoms into the ionization volume. A trivial way of increasing
the ionization speed and the spatial resolution not shown in the
simulation is simultaneously decreasing the beam radius of both
laser beams. Here all numbers are given for trapped and cooled
$^{87}$Rb atoms, however similar schemes can be found also for
other species \cite{Hurst79}.

\section{Ionization on the atomchip}

\begin{figure}[tbp]
\includegraphics{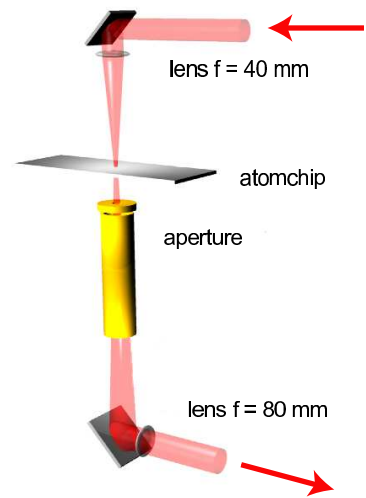}
\caption{(Color online) Experimental setup. The laser beams are injected from
outside the vacuum chamber and reflected by a mirror. A lens with
a focal length of 4 cm focusses the light through a hole with a
diameter of 150 $\protect\mu $m in the atom chip onto the trapping
region of the atoms. Behind the trap the light passes an aperture
to minimize stray light. A mirror and a second collimation lens
guides the beams out of the chamber.} \label{setup}
\end{figure}

The atomchip used in this experiment is described in detail
elsewhere \cite{Guenther05a}. It carries a conductor geometry
which allows for transporting condensates or thermal atoms to
different locations at the chip in a controlled way. At these
locations the atoms can be brought in contact with specialized
trapping geometries such as narrow wave guides, double well
potentials or periodic lattice potentials \cite{Guenther05} each
individually addressable by the transport system. Instead of
actively transporting the atoms it is also possible to couple
atoms into a waveguide and let them propagate ballistically
\cite{Fortagh03}. To this scenario we have added a region where
the atoms are ionized by the lasers. A sketch of the setup is
shown in Figure \ref{setup}.
Both laser beams are overlapped outside the vacuum chamber and
focussed on to the chip with a lens of $f=40$ mm focal length
located inside the vacuum chamber. The beams with a waist of
$w_0=30~\mu $m are aligned perpendicular to the surface of the atom
chip and pass the chip through a small hole with a diameter of 150
$\mu $m. With a Rayleigh length of the beams of 3.6 mm and 2.9 mm,
respectively, the intensity variation with the distances of the
atoms to the chip surface can be neglected. The same holds for the
slightly different positions of the foci due to the chromatic
abberation of the lens. By passing the beams through a small hole
in the chip instead of an alignment parallel to the chip surface
stray light is minimized. In addition the atoms can be ionized
very near to the surface, where narrow trapping potentials can be
realized with structures on a spatial scale of $\mu$m and below.
Behind the chip the laser beams leave the chamber through an
antireflection coated window.

Also conceivable is guiding the laser beams through an optical
fiber. However, at the end facette of the fiber the
beams rapidly diverge which makes it hard to control stray light.

\section{Experimental observations}

In the experiment $5\cdot 10^{5}$ $^{87}$Rb atoms in the
$F=2,m_{\mathrm{F}}=2$ state are prepared in a magnetic micro trap
at a temperature of 5 $\mu$K. The power of the diode and the
fiber laser beams are 8 mW and 2 W, respectively. The wavelength
of the diode laser is stabilized to the $5S_{1/2},F=2\rightarrow
5D_{5/2},F=4$ two photon transition recorded with $ ^{87}$Rb in a
vapor cell \cite{Kraft2005}.

The atoms are trapped below the hole in the chip at a distance of
350 $\mu $m from the chip surface. The length of the atomic cloud
in the trap amounts to approximately 1 mm. Typical trapping
frequencies are $2\pi \cdot 16$ Hz in axial direction and $2\pi
\cdot 160$ Hz in radial direction. Ionization is initialized by exposing the atoms to the laser fields for 10 ms. The remaining atoms are detected by absorption imaging after 1 ms of ballistic expansion. 

\subsection{Dipole forces and ionization}

\begin{figure}[tbp]
\includegraphics[width=5cm]{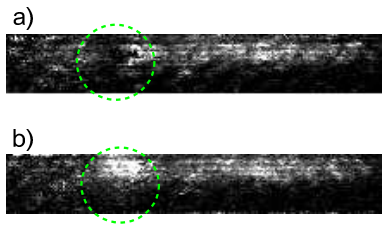}
\caption{(Color online) Absorption image of the atomic cloud. The cloud is
exposed (a) to the diode laser at a wavelength of 778 nm and with 8
mW. (b) to the fiber laser at 1080 nm and with 2 W. The dashed
circle indicates the area of exposure} \label{absorption}
\end{figure}

Fig. \ref{absorption} shows
absorption images for the cases that (a) only the diode laser or (b)
only the fiber laser was activated. In the first case the
density distribution is depleted at the position where the laser
illuminates the cloud. With only the fiber laser activated an
increase of the density is observed. Both effects are due to the
dipole forces of the laser beams.

\begin{figure}[tbp]
\includegraphics[width=8cm]{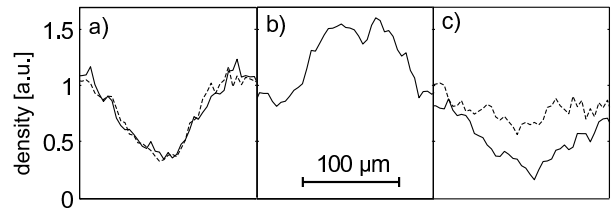}
\caption{Integrated density profiles. In (a) and (b) only one laser
was turned on. As in Fig. \ref{absorption} the dipole forces lead
to an increase of the local density. In (c) both lasers are turned
on. With the diode laser tuned out of resonance both dipole forces
cancel (dashed line). In resonance additional losses due to
ionization appear (solid line). The integrated density is
normalized to the unperturbed density distribution.\label{diff}}
\end{figure}

To investigate this in more detail Fig. \ref{diff}
shows the integrated atomic density profile at the position of the laser
beams. In (a) the diode laser was turned on for 10 ms and tuned in
resonance with the two photon transition $5S_{1/2}F=2\rightarrow
5D_{5/2}F=4$ of $^{87}$Rb (solid line). The dashed line shows the
density with the diode laser tuned 576 MHz to the red of this
transition. For both frequency settings the dipole potential due
to the diode laser is almost identical and depletes the density as
expected. There is no sign of additional losses or heating while
the diode laser is tuned to resonance. This proves that resonant
photon scattering does not affect the density on an observable
level. Furthermore it shows that the power of the 778 nm diode laser is to low to cause significant ionisation itself.

Fig. (b) shows the integrated density distribution for the analog
situation with only the fiber laser turned on for 10 ms. The laser
introduces an attractive potential and thus an increase of the
local density. Due to the far detuning of the laser no additional
heating is expected.

In Fig. (c) both lasers are turned on at the same time. Again the
diode laser was tuned 576 MHz red to the two photon resonance
(dashed line). The two dipole forces cancel and leave the density
profile almost unperturbed. With diode laser tuned in resonance
with the two photon transition (solid line in Fig. (c)) the density
is significantly depleted. This can not be explained with the
dipole potentials which are identical in and out off resonance.
Since ions can not be detected by the imaging laser, the additional losses are due to photo ionization which thus can
be directly observed spatially resolved in the depletion of the
density profile.

\subsection{Resonant ionization Spectrum}

\begin{figure}[tbp]
\includegraphics[width=8cm]{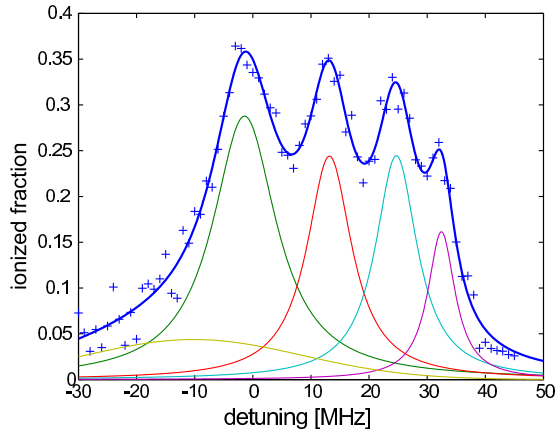}
\caption{(Color online) Resonant ionization spectrum. Tuning the diode laser
across the resonances of the 5S $\rightarrow $ 5D transition leads
to losses proportional to the number of ionized
atoms.\label{spectra}}
\end{figure}

Ionization spectra can be recorded by taking images for different
detunings of the diode laser. Here, the laser beams were turned on
for one millisecond and the atoms were imaged after one
millisecond of ballistic expansion. From the losses of atoms in a
region of interest around the focus of the laser beams the ionized
fraction can be determined (Figure \ref{spectra}).
Zero detuning corresponds to the center frequency of the $
5S_{1/2},F=2\rightarrow 5D_{5/2},F=4$ transition as observed in a
reference vapor cell. The ionization data (crosses) can be fitted
to a combination of four Lorentzian profiles plus one Gaussian
profile which models the background (solid line). The four maxima
match the expected positions of the transitions $F=2\rightarrow
F=4\dots 1$, however, width and relative heights deviate from the
expected spectrum of free and unperturbed atoms. Both deviations
are partly due to a common origin. The ionization was performed in
a magnetic trap with an non vanishing magnetic field of about 1
Gauss which is sufficiently strong to split the magnetic sub
states by some MHz \cite{Baluschev00}. The Zeeman splitting thus leads to a
broadening as well as to a reduction of the peak hight. As the
number of sub states increases with increasing $F$ this effect is
most dominant for the $F=4$ state with its 9 magnetic sub states.
Additional broadening may occur due to saturation of the two
photon transition and for large ionization rates which reduce the
lifetime of the 5D state.

\section{Conclusion}

In summary, we have reported the spatially resolved direct
observation of photo ionization of ultracold $^{87}$Rb atoms on an
atomchip. The atoms are ionized by absorbing three photons from
two continuous laser beams. The ionization is detected by
observing light induced losses of trapped atoms. Ionization
spectra are recorded by scanning across the resonance of the two
photon transition from the 5S ground state to the 5D excited
state. The combination of two lasers allow for compensating
unwanted dipole potentials which may repel the atoms from the
ionization region. The setup avoids stray light and is suitable
for single atom detection in combination with a sensitive ion
detector. Furthermore it may form the starting point for studying
ion atom mixtures in a combined micro trap for charged and neutral
particles.


\begin{acknowledgments}
We gratefully acknowledge support by the Landesstiftung Baden-W\"{u}%
rttemberg and by the Deutsche Forschungsgemeinschaft.
\end{acknowledgments}


%\bibliography{bibfile2}

\end{document}